# Decision Models for Workforce and Technology Planning in Services

**Gang Li**

Management Department

Bentley University

Waltham, MA, USA, 02452

gli@bentley.edu

**Joy M. Field**

Carroll School of Management

Boston College

Chestnut Hill, MA, USA, 02467

fieldjo@bc.edu

**Hongxun Jiang, Tian He, Youming Pang**

School of Information

Renmin University of China

Beijing, China, 100872

Abstract

Today's service companies operate in a technology-oriented and knowledge-intensive environment while recruiting and training individuals from an increasingly diverse population. One of the resulting challenges is ensuring strategic alignment between their two key resources – technology and workforce – through the resource planning and allocation processes.  The traditional hierarchical decision approach to resource planning and allocation considers only technology planning as a strategic-level decision, with workforce recruiting and training planning as a subsequent tactical-level decision. However, two other decision approaches – joint and integrated - elevate workforce planning to the same strategic level as technology planning.  Thus, we investigate the impact of strategically aligning technology and workforce decisions through the comparison of joint and integrated models to each other and to a baseline hierarchical model in terms of total cost.  Numerical experiments are conducted to characterize key features of solutions provided by these approaches under conditions typically found in this type of service company. Our results show that the integrated model is lowest cost across all conditions.  This is because the integrated approach maintains a small but skilled workforce that can operate new and more advanced technology with higher capacity.  However, the cost performance of the joint model is very close to the integrated model under a number of conditions and is easier to implement computationally and managerially, making it a good choice in many environments.  Managerial insights derived from this study can serve as a valuable guide for choosing the proper decision approach for technology-oriented and knowledge-intensive service companies.

# 1. Introduction

Many of today's service companies are both technology-oriented and knowledge-intensive (EUROSTAT, 2011; NCSES, 2012). Examples of such services include software development, information analytics, customer support, and product and service design. These types of services require strategic alignment between their two key resources – technology and workforce – throughout the organization, in order to provide outstanding service to customers (Davis and Heineke, 2012). This even applies to many original equipment manufacturers (OEMs) that are increasingly relying on after-sales services as a key source of revenue. In these companies, "the maintenance organizations [are] larger, more professional, and more complex" (Colen and Lambrecht, 2012: p.76), requiring a highly skilled and well-paid workforce. Two of the great challenges in doing this however, are "managing the workforce in a more technologically-oriented environment, and recruiting and training individuals from an increasingly diverse population" (Davis and Heineke, 2012: p.369). Motivated by these challenges, this paper investigates how the resource allocation and planning process can contribute to this necessary alignment between technology and workforce and how choosing the proper decision model can improve service system performance.

Service companies often face volatile demands and must make resource planning decisions well in advance in order to satisfy the expected demands with high service levels. Previous studies have provided various tools to incorporate the demand uncertainty and the service level requirement into the determination of the time-varying capacity levels in future periods (e.g. Hall, 1991; Gans et al., 2003; Green et al., 2007; Whitt, 2007). However, when operating in a technology-oriented and knowledge-intensive service environment that draws from a workforce pool with diverse skills, ensuring the necessary alignment of technology and employees to satisfy these time-varying capacity levels remains a challenge. Further complicating the resource allocation and planning process, the mix of older and newer technologies available to companies often require different employee skills. For example, while more advanced technology (e.g. better hardware and software for analyzing "big data") provides a higher capacity, this advantage is offset by the fact that the technology is often more expensive and requires specially trained employees to operate it. Conversely, older technology is often cheaper but less efficient. Determining the right types and quantities of technology to purchase, as well as the right time to purchase, are critical decisions for a company in securing the capacity sufficient to meet demand at all times. Meanwhile, different types of technology require different skill sets; only employees equipped with the appropriate skill set can operate a particular technology. Therefore, service companies must make workforce decisions to hire and/or train employees for the desired skill sets.

A company builds a portfolio of the employee skill sets in three different ways: by hiring new employees with the required skills; by training current employees; or by hiring and training, which is to first hire new and unskilled employees and then train them to the desired skills. To distinguish the latter two, we call the training of current employees as cross-training, and the training of new and unskilled employees as initial-training. Hiring new employees who hold the exact required skills brings immediate benefits to the company, although the cost associated with the new hiring can be expensive (Bidwell, 2011). Alternatively, cross-training current employees or initial-training new employees to obtain required skills may cost a company less money because the fixed costs associated with these decisions are often lower than the hiring cost of recruiting new and appropriately skilled employees (Park et al., 2011; Othman et al., 2012). However, training takes time and an employee in training cannot provide services to customers.

We develop mathematical models to support three resource planning and allocation decision approaches – hierarchical, joint, and integrated – in technology-oriented and knowledge-intensive service environments where service demand must be satisfied in the same period. With the hierarchical approach, technology planning is considered to be strategic, but workforce planning is tactical. The joint and integrated approaches, on the other hand, consider both technology and workforce planning to be strategic. Thus, to investigate the impact of strategically aligning technology and workforce decisions, we compare the joint and integrated models to each other and to a baseline hierarchical model in terms of total cost. In the models we consider multiple options for hiring and training to acquire and develop necessary employee skills.

We conduct numerical experiments to compare key features of solutions provided by the three approaches under different demand and supply scenarios. As a baseline, the hierarchical approach always has the highest cost but also achieves the highest resource utilization. The integrated approach always has the lowest cost but has the highest technology cost and lowest technology utilization. However, the cost performance of the joint model is very close to the integrated model under a number of conditions and is easier to implement computationally and managerially, making it a good choice in many environments. Thus, we identify conditions where the joint decision approach is a good substitute for the integrated decision approach. We present detailed explanations why different decision approaches lead to these different results and illustrate the critical role of cross-training in reducing the total cost.

Our work also provides a timely and quantitative analysis of the unbalanced spending of contemporary businesses on labor and capital. Usually, capital and labor are complementary: a business that buys a new truck often hires a new driver, too. According to the United States Department of Commerce, however, business spending for employees has grown only 2 percent, but equipment and software spending has increased 26 percent as the US economy recovers from the 2008-2009 recession (Rampell, 2011). It seems that companies are spending more on technology, not new workers, as reported by *The New York Times* (Rampell, 2011). The insights drawn from the integrated model explain why this imbalance happens, as seen when companies use training to offset high and increasing labor costs and take advantage of the decrease in technology costs.

The rest of the paper is organized as follows. Section 2 reviews the related literature. Section 3 first analyzes the key components associated with the decision process and then develops a baseline hierarchical decision model and the joint and integrated models that are the focus of our study. For each decision model, we discuss time-efficient algorithms to solve it optimally. Through numerical experiments, Section 4 characterizes key features of the optimal solutions provided by the three approaches and compares the cost performance of the joint and integrated models to each other and to the baseline hierarchical model. Insights, recommendations, and limitations are addressed in Section 5.

## 2. Literature Review

The design and management of the underlying processes for technology-oriented and knowledge-intensive services tend to be complex (Marjanovic and Freeze, 2012) for a number of reasons and this complexity impacts the performance of resource planning and allocation decision approaches. First, in these types of services, as with services in general, demand is often unpredictable, yet customers are increasingly expecting higher service levels (Easton, 2014; Van den Schrieck et al., 2014).

Second, while service providers leverage technology to fulfill the service need, their required skills, knowledge, and judgment are critical components that drive service performance (Marjanovic and Freeze, 2012; von Nordenflycht, 2010; Bettencourt et al., 2002). Different types of people differ in their sets of skills (e.g. technical, managerial, business, interpersonal), experiences, personalities, and motivations; these features directly impact workers' performance (Hunter et al., 1990; Hurtz and Donovan, 2000; Kanfer and Ackerman, 2000; Penney et al., 2011). Researchers have modeled worker differences into the workforce management framework for general purposes (e.g., Billionnet, 1999; Norman et al., 2002; Thompson and Goodale, 2006; Othman et al., 2012), or for special applications such as information system management (Lee et al., 1995; Byrd and Turner, 2001; Palshikat et al., 2011) and call-center scheduling (Wallace and Whitt, 2005; Avramidis et al., 2009; Cordone et al., 2011).

Further complicating matters, these service processes need employees with specific skills that can be acquired in the labor market but can also require ongoing training and cross-training to use both various existing technologies and also new and better technologies over time (Bidwell, 2011). Training, as an important way to learn, has long been acknowledged by practitioners and researchers to be effective in improving service quality and increasing productivity (MacDuffie, 1995; Chuang and Liao, 2010; Jiang et al., 2012). Many theoretical frameworks have been developed to incorporate the training option into the heterogeneous workforce setting. Gans and Zhou (2002) assumed discrete skill levels for employees, and used training to transfer employees from a lower level of skills to a higher level. Hopp and Van Oyen (2004) proposed frameworks to design effective cross-training architectures from both strategic and tactical perspectives in order to produce agility. Akşin (2006) built a linear programming model to derive optimal decisions on hiring, firing, and training across various service demand patterns.

While cross-training has important workforce development benefits and helps to pool employee resources to mitigate the effects of demand uncertainty (Easton, 2011; 2014), these benefits may have limits. According to Hopp and Van Oyen (2004), training inefficiencies (i.e., high costs of skill acquisition and high skill variation) and switching inefficiencies (e.g. if an employee needs to travel a long distance to perform another task), need to be considered when determining the best amount of cross-training. In addition, with multiple tasks, flexible workers do not gain as much experience as their specialized counterparts in a particular task, and this lack of experience negatively impacts service quality (Pinker and Shumsky, 2000). Overall, "Although cross-training increases server flexibility and improves responsiveness, it also increases the service costs and may reduce service efficiency" (Chakravarthy and Agnihothri, 2005: p.218).

All the existing work on heterogeneous workforce management has implicitly assumed that physical capacity (i.e., technology availability) is unlimited, which is often not borne out in reality. In addition, when considering technology and workforce planning and allocation together, employee capacity depends on the type of technology assigned to the employee. For example, an employee will have a higher capacity if assigned to a better technology. In our model we incorporate the idea that the capacity of the workforce depends on both the employees' skill sets and the capabilities of the assigned technology.

Thus, given the critical roles that both technology and workforce play in technology-oriented and knowledge-intensive services, the alignment of technology and workforce is a key strategic consideration when developing models for resource planning and allocation (Ray et al. 2005; Gaimon et al., 2011; Hopkins and Brokaw, 2011). However, in the manufacturing sector, capacity planning decisions have traditionally been viewed as a strategic decision process made by the top management of a company based mainly on long-term capacity level and utilization of physical (such as, technology) resources. The optimal workforce level in the company, on the other hand, was treated as a secondary decision issue made by middle management, and often considered in the aggregate planning decision process along with the detailed production rate decision (Hopp and Spearman, 2008). A hierarchical approach has been recommended for decades to solve the basic resource planning and allocation decision problem (Hax and Meal, 1975; Bitran and Tirupati, 1993). This approach divides the decision process into three steps: Starting from a strategic level, the company first determines the optimal technology purchasing plan to ensure there is sufficient capacity to satisfy the future requirements. Once the time-varying capacity requirements are satisfied and the technology becomes available, the hierarchical approach then considers a tactical-level workforce planning decision that hires or trains employees to use the technology. Finally, given available technology and employees, an operational-level resource assignment decision is used to match technology with the proper employees to best serve customer demand. However, a key limitation of the hierarchical approach for the types of services considered here it that workforce recruiting and training is modeled as a tactical, not a strategic, decision.

Similar to the hierarchical decision approach dominating capacity and technology management, decisions in workforce management have also long been made in a hierarchical approach (Abernathy et al., 1973; Alfares, 2004), which includes policy decisions (such as scheduling flexibility and skill patterns), staffing decisions (such as staff size and employee training), and detailed labor scheduling and allocation decisions. Meanwhile, various models have been proposed to make joint workforce decisions, such as joint staffing and cross-training decisions (Kao and Tung, 1981), and joint scheduling and allocation decisions (Warner and Prawda, 1972). Easton (2011) provides a good survey on these joint decision models. In recent works, Wirojanagud et al. (2007) and Fowler et al. (2008) develop staffing models that that capture the hiring, firing, and cross-training decisions and propose heuristics to solve the models. Campbell (2011) and Easton (2011; 2014) study joint scheduling and allocation models for cross-trained workers. In their work, a two-stage stochastic program is proposed, in which the first stage determines the workforce size and generates an initial schedule based on the expected demand and the second stage then adjusts the schedule based on the realized demand.

Due to the close interactions between technology and workforce and the need for strategic alignment, it is obviously beneficial to have a joint technology-workforce plan. Two types of decision approaches – joint and integrated – elevate workforce planning to the same strategic level as technology planning. A service company may adopt a joint decision approach that would first pair technology with the best matched worker, and then make the joint resource planning decision and the subsequent assignment decision. An integrated decision approach, however, makes all these decisions simultaneously in a single step. Figure 1 compares the hierarchical, joint, and integrated decision approaches in terms of the decision type (i.e., strategic, tactical, and operational) for each of the planning and assignment decisions.

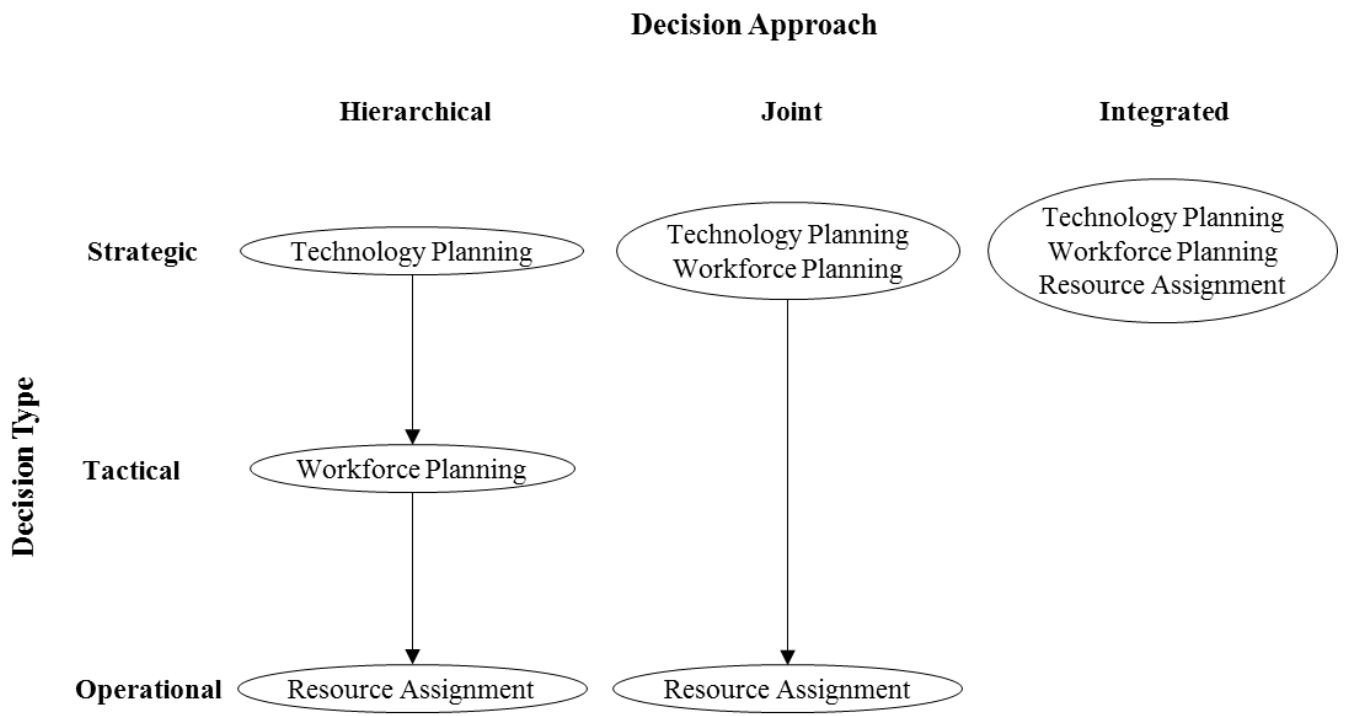

Figure 1. Comparison of Decision Models for Resource Planning and Allocation

To the best of our knowledge, Behnezhad and Khoshnevis (1988) wrote the first paper proposing the idea of integrating capacity, including technology planning and workforce planning decisions. Behnezhad and Khoshnevis (1996) developed a more comprehensive framework, in which they presented a mathematical model including hiring/firing decisions for the workforce, procurement decisions for machines, and production decisions that assign workers to machines. Their model, however, still assumed homogeneity of both machines and workers. Because machines and workers were essentially identical in Behnezhad and Khoshnevis (1996), there was no need to match different types of machines with different types of workers. Their setting also precluded the need for workforce training. He and Jiang (2011) also developed an integrated model that incorporates multiple resources, but did not consider demand uncertainty, equipment replacement, or workforce termination.

We present more general models than Behnezhad and Khoshnevis (1996) and He and Jiang (2011) to support the integrated decision approach. In addition, we develop a joint decision model for comparison to the integrated approach. Unlike Behnezhad and Khoshnevis (1996), however, resources in our model are *heterogeneous*, which better reflects the service environment under consideration. An optimal service plan must select and assign different types of technology with the proper types of workers in order to satisfy time-varying capacity requirements at a minimum cost. Meanwhile, workers can add skills (i.e., change their types) by cross-training during the planning horizon. In summary, our paper extends previous studies to develop a more general integrated framework, in which integration considers not only workforce decisions, but also technology and capacity decisions. And within the workforce decision category, we capture the dynamics of workforce development by modeling the training time, cost, and potential training paths, which is a generalization to Wirojanagud et al. (2007). Meanwhile, aimed for high level decision support, the decision models in our study do not have the detailed scheduling constraints as in Campbell (2011) and Easton (2011, 2014). Overall, effectively solving the decision problem and promoting strategic alignment among resources entails breaking the traditional boundaries of strategic capacity planning, tactical workforce planning, and operational resource assignments.

## 3. Models to Support Different Decision Approaches

### 3.1. Decisions in a technology-oriented service environment with a diverse workforce

The models we build in this paper support workforce recruiting/training/firing, technology purchasing/discarding, and their assignments for a service firm in a technology-oriented environment that requires diverse workforce skills. In such an environment, technology varies in technical specifications, while the workforce varies as to whether it qualifies for these specifications. In addition, the service work is typically knowledge-intensive with employee wages that are high relative to the cost of technology. For example, to analyze SAP transactional business operations in real time, a high tech company may purchase an IBM System x3690 X5 Workload Optimized Solution for SAP HANA (IBM-SAP, for short). Part of the technical specifications of the solution includes the IBM x3690 X5 Server, Linux Operating System, IBM General Parallel File System, and SAP HANA (IBM Server, 2013). A qualified employee has to master all these technical specifications. Moreover, a qualified employee may master other technical specifications in addition to what the IBM-SAP solution requires, such as knowing how to run a HP server. Thus, the company must distinguish the two types of qualified employees, one only knowing the IBM-SAP solution and the other knowing both IBM-SAP and HP, and choose one of them to work with the IBM-SAP solution.

We define each technical specification as a *skill* and denote *K* as the set of all technical specifications. Both technology and workforce are distinguished from each other by the skills they require or hold. Denote the set of available technology types as *I* and the set of available employee types as *J*. For each type of technology, $i \in I$, the skills required to operate are a subset of *K*, denoted as $KE_i$, $KE_i \subseteq K$. For each type of employee, $j \in J$, the skills held are also a subset of *K*, denoted as $KW_j$, $KW_j \subseteq K$. Thus, if and only if $KE_i \subseteq KW_j$ holds for a technology-employee type pair (*i*, *j*), a type *j* employee is *qualified* to operate a piece of type *i* technology (e.g. an individual license for a software program or piece of hardware). Denote the set of all qualified technology-employee type pairs (*i*, *j*) as $\Psi$. Once operated, a piece of type *i* technology provides a capacity $c_i$ to customer demand. A type *j* employee can change into a new type *j'* by receiving training. However, such a change is *feasible* if and only if $KE_j \subset KE_{j'}$. Denote the set of all feasible training pairs (*j*, *j'*) as $\Omega$. For any $(j, j') \in \Omega$, the training takes $l_{jj'}$ periods of time. We assume an employee does not serve customers during the training periods.

At the beginning of a planning horizon (e.g. at the beginning of each fiscal year), the company first determines desired capacity levels in future periods based on time-varying and stochastic demands and specified service requirements. The company then makes an effective resource plan, given the resources it already owns, in order to satisfy these capacity requirements in the planning horizon. Our decision models focus on the determination of the optimal resource plans by assuming that the desired capacity levels are specified *a priori*. Denote $BX_i$, $i \in I$, as the set of technologies (i.e., hardware and software) owned by the company, and $BY_j$, $j \in J$, as the set of employees working in the company at the beginning of the planning horizon. We assume the initial resources must be *balanced*, that is, each piece of existing technology must have a current employee to operate it, $\sum_{j \in J} BY_j = \sum_{i \in I} BX_i$. The planning horizon consists of $T$ time periods, numbered from 1 to $T$. Let $t$ ($1 \leq t \leq T$) be the index of time period and $d_t$ be the desired capacity levels in period $t$. The company needs to decide in each period whether to purchase new or discard existing technology, in which quantity and of which type. The company also needs to decide in each period whether to hire new or fire current employees, in which quantity and of which type, as well as whether to train current employees in certain quantities and make determinations from type to type. Meanwhile, the company has to decide how to assign qualified employees to technology in order to reach the desired capacity level in each period. Corresponding to the above decisions, we define six types of variables in each period $t$: let $x_{it}$ be the amount of type $i$ technology purchased, $v_{it}$ be the amount of type $i$ technology discarded, $y_{jt}$ be the number of type $j$ employees hired, $w_{jt}$ be the number of type $j$ employees fired, $u_{jj't}$ be the number of type $j$ employees starting to be trained to be type $j'$ employees, and $z_{ijt}$ be the amount of type $i$ technology assigned to type $j$ employees. Each decision incurs a certain amount of cost. Corresponding to each decision, we define the following six types of costs in each period $t$: $p_{it}$ is the unit technology purchasing and maintenance cost, $s_{it}$ is the unit technology discarding cost, $h_{jt}$ is the initial unit workforce cost, $f_{jt}$ is the unit workforce firing cost, $r_{jj't}$ is the unit training and incremental workforce cost (based on the value of added skills), and $m_{ijt}$ is the unit assignment cost.

A salient feature of our modeling approach is that all decisions and costs in our models are time-dependent due to the fact that a decision impacts both the current period as well as the future. Take a technology purchasing decision as an example. Once a company decides to purchase a piece of type $i$ technology in time $t$, the company will pay a one-time purchasing cost in period $t$. After that, the company will also pay a maintenance cost in each of the future periods. Thus, the technology purchasing decision in the planning horizon consists of both a one-time cost and recurring costs. We define

$$p_{it} = PC_i + MS_i \sum_{t'=t}^{t'=T} \gamma^{t'-t}, \forall i \in I, 1 \leq t \leq T \qquad (1)$$

where $PC_i$ is the one-time purchasing cost, $MS_i$ is the maintenance cost in each period, and $\gamma$ is the discount factor, $0 < \gamma < 1$. Consequently, when the company decides to discard a piece of existing technology, it incurs both a one-time cost and recurring "benefits" as the formulation shown below:

$$s_{it} = DC_i - MS_i \sum_{t'=t}^{t'=T} \gamma^{t'-t}, \forall i \in I, 1 \leq t \leq T \qquad (2)$$

where $DC_i$ is the one-time discarding cost. Equation (2) implies that the company will save maintenance costs in future periods by discarding the technology now. Thus, $s_{it}$ could be a negative cost, i.e., a profit for the company. We use a similar way to define the workforce-related costs, in which the recurring cost is the base salary, i.e., the salary independent of the tasks assigned, paid to an employee in each period. Note that in the training decision, when an employee changes their type from $j$ to $j'$, the recurring cost is the difference in base salaries between $j'$ and $j$. The task-dependent salary is counted in the assignment cost $m_{ijt}$. By defining time-dependent decisions and costs, we do not need to explicitly define recurring costs associated with each decision. Such a feature makes the models presented in the next three subsections concise and easy to solve. Our models support three different decision approaches: the baseline *H*ierarchical approach, where technology planning is strategic but workforce planning is tactical, and the *J*oint and *I*ntegrated approaches, where technology and workforce decisions are strategically aligned. Each approach provides its answers to three key decision problems: *W*orkforce *P*lanning, *T*echnology *P*lanning, and *R*esource *A*ssignment. To simplify notations, we use *H*, *J*, and *I* to represent the three approaches, and *WP*, *TP*, and *RA* to represent the three decisions, respectively. *TC* denotes the *T*otal *C*ost resulting from a decision approach. Table 1 summarizes all notations addressed above.

| Sets: | Subsets: |
|---|---|
| <ul><li>$K$: set of technique skills</li><li>$I$: set of available technology types</li><li>$J$: set of available employee types</li></ul> | <ul><li>$KE_i \subseteq K$: skills required to operate type $i$ technology</li><li>$KW_j \subseteq K$: skills held by type $j$ employee</li><li>$\Psi = \{(i,j) \mid i \in I, j \in J, KE_i \subseteq KW_j\}$: *qualified* technology-employee type pairs</li><li>$\Omega = \{(j,j') \mid j, j' \in J, KE_j \subset KW_{j'}\}$: *feasible* workforce training pairs</li></ul> |
| **Parameters:** | |
| <ul><li>$c_i$: unit capacity of type $i$ technology</li><li>$T$: total planning periods</li><li>$\gamma$: discount factor, $0 < \gamma < 1$</li><li>$BX_i$: Initial set of technology ($i \in I$)</li></ul> | <ul><li>$l_{jj'}$: training time from type $j$ to type $j'$</li><li>$t$: index of the planning periods ($1 \leq t \leq T$)</li><li>$d_t$: desired capacity level at period $t$ ($1 \leq t \leq T$)</li><li>$BY_j$: Initial set of employees ($j \in J$)</li></ul> |
| **Decision Variables and Corresponding Costs** | |
| <ul><li>$x_{it}$: technology purchasing variable</li><li>$v_{it}$: technology discarding variable</li><li>$y_{jt}$: workforce hiring variable</li><li>$w_{jt}$: workforce firing variable</li><li>$u_{jj't}$: workforce training variable</li><li>$z_{ijt}$: technology-workforce assignment variable</li></ul> | <ul><li>$p_{it}$: unit technology purchasing and maintenance cost</li><li>$s_{it}$: unit technology discarding cost</li><li>$h_{jt}$: unit initial workforce cost</li><li>$f_{jt}$: unit workforce firing cost</li><li>$r_{jj't}$: unit training and incremental workforce cost</li><li>$m_{ijt}$: unit technology-workforce assignment cost</li></ul> |
| **Abbreviations for Decision Approaches and Key Decisions** | |
| <ul><li>Abbr. for decision approaches: *H* for Hierarchical; *J* for Joint; *I* for Integrated</li><li>Abbr. for key decisions: *WP* for workforce planning; *TP* for technology planning; *RA* for resource assignment; *TC* for total cost.</li></ul> | |

Table 1. Summary of Notations

## 3.2. Models and Solution Algorithms for the Hierarchical Decision Approach

The hierarchical decision includes three sequential steps: technology planning, workforce planning, and resource assignment; each subsequent step is based on the optimal results of the previous step. We present the decision models for each planning step in Appendix A.

The total cost resulting from the hierarchical decision approach, denoted by *HTC*, is equal to:

$$HTC = HEP(x_{it}^*, v_{it}^*) + HWP(y_{jt}^*, w_{jt}^*, u_{jj't}^* \mid x_{it}^*, v_{it}^*) + HRA(z_{ijt}^* \mid y_{jt}^*, w_{jt}^*, u_{jj't}^*, x_{it}^*, v_{it}^*) \quad (3)$$

where $HEP(x_{it}^*, v_{it}^*)$ is the optimal technology planning cost at step 1, $HWP(y_{jt}^*, w_{jt}^*, u_{jj't}^* \mid x_{it}^*, v_{it}^*)$ is the optimal workforce planning cost at step 2 given the technology planning decisions, and $HRA(z_{ijt}^* \mid y_{jt}^*, w_{jt}^*, u_{jj't}^*, x_{it}^*, v_{it}^*)$ is the optimal resource assignment at step 3 given the resource planning decisions at the previous two steps.

**Proposition 1**: If the technology discarding cost is non-negative, i.e., $s_{it} \geq 0$, the optimal technology planning problem can be solved as a *Time-Dependent Knapsack Problem*.

**Proof**. Please see Appendix B.

**Proposition 2**: If the technology discarding cost is non-negative, i.e., $s_{it} \geq 0$, the optimal workforce planning problem can be solved as a series of *Time-Constrained Shortest Path Problems*.

**Proof**. Please see Appendix C.

**Proposition 3**: The optimal resource assignment problem can be solved as a series of *Bounded Knapsack Problems*.

**Proof**. Please see Appendix D.

Propositions 1-3 suggest that if discarding technology incurs positive costs, every step in the hierarchical decision can be solved as one or a series of classical problems. Note that although all these problems are NP-complete, there are pseudo-polynomial time algorithms to solve them efficiently. The existence of time-efficient algorithms is consistent with the conventional belief that a hierarchical decision approach is easy to use in practice. In the case of negative technology discarding cost, we can use the general Integer Programming algorithm to solve the resource planning decisions steps 1 and 2.

### 3.3. Models and Solution Algorithms for the Joint Decision Approach

The joint decision approach combines technology planning and workforce planning together in order to manage the two strategic resources in a coordinated way. This approach *a priori* matches each type of technology with a most preferred type of worker. Given the set of matched technology-workforce pairs, a joint decision is made to determine the technology purchasing/discarding, and the workforce hiring/firing/training simultaneously. Finally, an assignment decision is made. Three decision models, as described below, are needed to support the decision procedure.

**Step 0. Pre-processing: Define the set of preferred technology-workforce pairs**

Assuming a piece of type *i* technology is purchased in time *t*, a worker with the most preferred type, denoted by $j_{it}$, must also be hired. The preference is based on minimum cost of obtaining a qualified worker for a given piece of technology. The following model shows how to find the preferred worker type, given each $i \in I$ and $1 \leq t \leq T$. We denote this problem as Workforce Selection Problem, **WSP** for short.

**WSP**: 
$$g_{j_{it}} = min \frac{1}{\gamma^{t-1}} \sum_{t'=1}^{t} \gamma^{t'-1} \left[ \sum_{j \in J} h_{jt'} y_{jt'} + \sum_{j \in J} \sum_{\{j'|(j,j') \in \Omega\}} r_{jj't'} u_{jj't'} \right] \quad (4)$$

s.t.
$$\sum_{\{j|(i,j) \in \Psi\}} Y_{jt} = 1 \quad (5)$$

$$Y_{jt'} = Y_{j,t'-1} + y_{jt'} + \sum_{\{j'|(j',j) \in \Omega, t''=t'-l_{j'j}>0\}} u_{j'jt''} - \sum_{\{j'|(j,j') \in \Omega\}} u_{jj't'}, \quad \forall j, t' < t \quad (6)$$

$y_{jt'}$, $u_{jj't'}$, and $Y_{jt'}$ non-negative integers  (7)

Model **WSP** minimizes the total cost of obtaining a most preferred employee, including initial workforce cost and training cost, if any, for a piece of purchased technology type *i* in time *t*. Constraint (5) ensures that one employee with the qualified type *j*, i.e., $(i, j) \in \Psi$, must be available in period *t*. A qualified employee can be obtained either by direct hiring or through a series of initial training sessions; constraint (6) shows all feasible options of obtaining such a qualified employee. Model **WSP** can be transformed into the shortest path problem on a network constructed as follows: The network has nodes at each $j$ ( $j \in J$ ) and $t'$ ( $1 \leq t' \leq t$ ) and two dummy nodes. A source dummy node connects to all other $(j, t')$ nodes with cost $\gamma^{t'-t} h_{jt'}$ and a destination dummy node is connected by all $(j, t)$ nodes satisfying $(i, j) \in \Psi$ with cost 0. Additionally, there is an arc from each node $(j', t'')$ to node $(j, t')$ with cost $\gamma^{t'-t} r_{jj't'}$ if $(j', j) \in \Omega$ and $t'-t'' = l_{j'j}$, and an arc from each node $(j, t')$ to node $(j, t'+1)$ ($1 \leq t' < t$) with a nominal cost. Thus, model **WSP** is equivalent to the shortest path (i.e., minimum cost path) problem from the dummy source node to the dummy destination node in an acyclic network as constructed above, which can be solved optimally by a dynamic programming algorithm in $O(|J|*T)$ (Ahuja et al., 1993).

**Step 1. Joint Decision of Technology Planning and Workforce Planning.**

**J_S1:** $\quad JEPWP(y^*_{jt}, w^*_{jt}, u^*_{jj't}, x^*_{it}, v^*_{it}) = \min \quad \sum_{t=1}^{T} \gamma^{t-1} \sum_{i \in I} [(p_{it} + g_{j_{it}}) x_{it} + (s_{it} + f_{j^*_i t}) v_{it}]$ (8)

$\quad$ s.t. $\quad \sum_{i \in I} c_i (BX_i + \sum_{t'=1}^{t} x_{it'} - \sum_{t'=1}^{t} v_{it'}) \geq d_t, \quad \forall 1 \leq t \leq T$ (9)

$\Delta x_{it}$ and $\Delta v_{it}$ are non-negative integers (10)

Note that in model **J_S1**, an employee is always paired with a piece of technology. If a piece of technology is purchased, a preferred employee is also hired; and if a piece of technology is discarded, the employee who operates it is also fired. Thus, we use only variables $x_{it}$ and $v_{it}$ to represent resource purchasing and discarding decisions. Note that the same type of technology may be paired with different types of preferred employees in different purchasing times. Thus, the types of employees a company can fire are dependent of the types of employees the company actually hired and cannot be determined *a priori*. To simplify the model, we define a nominal firing cost, $f_{j^*_i t} = \min\{f_{jt} \mid j \in \{j_{it'} \mid 1 \leq t' \leq t\}\}$, corresponding to each technology discarding decision with type $i$ and time $t$. Recall that $j_{it'}$ is the preferred employee type if a piece of type $i$ technology is purchased in time $t$. Thus, the set $\{j_{it'} \mid 1 \leq t' \leq t\}$ represents all possible employee types that can be paired with a piece of type $i$ technology until $t$. We use the minimum firing cost among employee types in the set as the nominal cost. Once model **J_S1** is solved and the actual employee hiring decisions are known, the nominal cost is then adjusted by the actual firing cost. Model **J_S1** has the same structure as model **H_S1**. Therefore, the conclusion of Proposition 1 is also applicable.

**Step 2. Resource Assignment.**

$$\textbf{J\_S2:} \quad JRA(z^*_{ijt} \mid y^*_{jt}, w^*_{jt}, u^*_{jj't}, x^*_{it}, v^*_{it}) = \min \sum_{t=1}^{T} \gamma^{t-1} \sum_{i \in I} m_{ij_it} z_{ij_it} \quad (11)$$

$$s.t. \quad \sum_{i \in I} c_i z_{ij_it} \geq d_t, \quad \forall t \quad (12)$$

$$z_{ij_it} \leq BX_i + \sum_{t'=1}^{t} x^*_{it'} - \sum_{t'=1}^{t} v^*_{it'} \quad \forall i, t \quad (13)$$

$$z_{ij_it} \text{ are non-negative integers} \quad (14)$$

Model **J_S2** has the same structure as model **H_S3**. Therefore, the conclusion of Proposition 3 also applies to it. The total cost resulting from a joint decision, denoted by *JTC*, is equal to:

$$JTC = JEPWP(y^*_{jt}, w^*_{jt}, u^*_{jj't}, x^*_{it}, v^*_{it}) + JRA(z^*_{ijt} \mid y^*_{jt}, w^*_{jt}, u^*_{jj't}, x^*_{it}, v^*_{it}) \quad (15)$$

By simultaneously determining type and quantity of both technology and workforce, the joint decision approach is expected to produce a better match of resources than the hierarchical decision approach, which only looks at the value of technology in the first step. By pairing each technology type with the preferred work type *a priori*, the joint decision approach also precludes the need of cross-training of employees during the planning horizon.

**3.4. Models and Solution Algorithms for the Integrated Decision Approach**

The integrated decision approach uses a single model to determine the optimal resource planning and assignment decisions. As with the joint decision model, it puts workforce planning at the same strategic level as technology planning. The model is presented as follows:

$$ITC = JEPWPRA(y^*_{jt}, w^*_{jt}, u^*_{jj't}, x^*_{it}, v^*_{it}, z^*_{ijt})$$

$$= min \sum_{t=1}^{T} \gamma^{t-1} \left[ \sum_{j \in J} h_{jt} y_{jt} + \sum_{j \in J} \sum_{\{j'|(j,j') \in \Omega\}} r_{jj't} u_{jj't} + \sum_{j \in J} f_{jt} w_{jt} + \sum_{i \in I} p_{it} x_{it} + \sum_{i \in I} s_{it} v_{it} + \sum_{i \in I} \sum_{\{j|(i,j) \in \Psi\}} m_{ijt} z_{ijt} \right] \quad (16)$$

s.t.

$$\sum_{i \in I} \sum_{\{j|(i,j) \in \Psi\}} c_i z_{ijt} \geq d_t \quad \forall 1 \leq t \leq T \quad (17)$$

$$\sum_{\{j|(i,j) \in \Psi\}} z_{ijt} \leq BX_i + \sum_{t'=1}^{t} x_{it'} - \sum_{t'=1}^{t} v_{it'} \quad \forall i, t \quad (18)$$

$$Y_{jt} = BY_j + y_{jt} + \sum_{\{j'|(j',j) \in \Omega, t'=t-l_{j'j}>0\}} u_{j'jt'} - \sum_{\{j'|(j,j') \in \Omega\}} u_{jj't} - w_{jt}, \quad t=1, \forall j \quad (19a)$$

$$Y_{jt} = Y_{j,t-1} + y_{jt} + \sum_{\{j'|(j',j) \in \Omega, t'=t-l_{j'j}>0\}} u_{j'jt'} - \sum_{\{j'|(j,j') \in \Omega\}} u_{jj't} - w_{jt}, \quad 2 \leq t \leq T, \forall j \quad (19b)$$

$$\sum_{\{i|(i,j) \in \Psi\}} z_{ijt} \leq Y_{jt} \quad \forall j, t \quad (20)$$

$x_{it}$, $y_{jt}$, $u_{jj't}$, $v_{it}$, $w_{jt}$, $z_{ijt}$, and $Y_{jt}$ are non-negative integers  (21)

The decision model minimizes the total cost of workforce hiring, firing, and training, technology purchasing and discarding, and resource assignment, subject to the capacity requirement constraint in (17) and the resource alignment constraints in (18)-(20). Without pre-specified resource pairs, the integrated model allows a piece of technology to be operated by different types of employees at different times, as well as an employee to be reassigned to another piece of technology after obtaining new skills through cross-training. Such new flexibilities could help decrease total cost. A question one might ask, however, is whether the benefit gained is worth the effort. Note that compared to the hierarchical and joint models, the integrated model requires solving a much bigger and more complex model, which does not have known time-efficient algorithms. In the case of non-negative technology discarding cost ($s_{it} \geq 0$) and workforce firing cost ($f_{jt} \geq 0$), we developed a very efficient genetic-based algorithm that is able to obtain optimal or close-to-optimal solutions in seconds for large scale problem instances (Li et al., 2015). For general problem instances, we rely on the Integer Programming algorithm and use IBM CPLEX to solve. However, when the model size continues to increase, CPLEX encounters difficulties in providing good quality solutions in an acceptable computational time. In addition to the computational challenge, the integrated decision approach calls for significant changes in management. The traditionally separated business units – human resource, supply chain, and customer service – have to join together to ensure that the integrated decision paradigm works. Therefore, it is not a trivial question to ask whether and to what extent the integrated model can outperform the joint model (assuming both models outperform the baseline hierarchical model) — or from where the main benefits of the integrated model will come. In the next section, we address these questions through extensive numerical experiments.

## 4. Numerical Experiments and Insights

The purpose of the numerical experiments is to provide practical guides for service managers to adopt the appropriate decision approach for their particular planning application. To this end, we will first evaluate the cost differences of optimal solutions that are provided by these three decision approaches in various environments, and then inspect in detail the key characteristics of each solution.

### 4.1. Design of Numerical Experiments

To simplify parameter configurations, we let skill set *K* be the same as technology set *I*, which implies that each type of technology requires a unique skill. Given a set *I*, an employee may operate none of the technology, one type, two types, and so on. Thus, the number of employee types is $2^{|I|}$. The planning horizon is set equal to 10 periods and the training time for obtaining one skill is generated randomly between 0 and 2 periods. The randomly generated desired capacity level has an average around 1000, from a range of 0 to 2000. Meanwhile, the unit capacity of each piece of technology is also randomly generated in a range of 200 to 1000. On the cost setting, we let the workforce-related cost generally be more expensive than the technology-related cost since, as previous studies pointed out (e.g., Aksin et al. 2007), the workforce-related cost is high in a typical knowledge-intensive service company. In Appendix E, we describe the parameter settings for each cost component and demonstrate an example instance.

To generate the initial technology and workforce sets, $BX_i$ and $BY_j$, properly, we assume that the values of *BX* and *BY* in each decision approach are the long-term outcomes for a company using the same decision approach in the past. Take the hierarchical decision approach as an example. Assuming the average capacity level in the past is *D*, we call the models of the hierarchical approach to solve a single-period problem, in which the desired capacity level is *D* and the initial sets are empty. We then let *BX* be the optimal solution of model **H_S1** and *BY* be the optimal solution of model **H_S2**. Similarly, we use decision models corresponding to the joint and the integrated approaches to derive the values of the initial sets for the two approaches, respectively. This method ensures that it is the decision approach – not the initial value setting – that drives the model performance. To link the past capacity requirements with the future, we further assume that *D* is equal to the average desired capacity level in the first period of the planning horizon.

The Integer Programming models used in the numerical experiments were implemented in Java using Concert Technology of IBM CPLEX 12. The experiments were conducted in a HP Compaq Elite 8300 desktop with Intel 3.2GHz i5-3470 CPU and 4GB memory.

### 4.2. Cost Comparison of Decision Approaches in Various Environments

Service companies often have to make decisions facing uncertain and volatile demands. The combination of technology-oriented and knowledge-intensive services, and the desire for strategic alignment between technology and the workforce, further complicate the decision procedure by offering a vast, often overwhelming, number of choices on the supply side. Moreover, various cost and time factors have impacts on the decisions as well. We therefore simulate service scenarios that include (a) volatile demands, (b) complex technology and workforce settings, and (c) different cost and time settings, in order to investigate how the total costs provided by joint and integrated approaches compare to each other and the baseline hierarchical approach when confronted with these challenges.

### 4.2.1. Decisions under Volatile Demands

Holding all else constant, the changes in demands directly vary with the values of the desired capacity level $d_t$, which are determined based on the demand and service uncertainties and on the service level requirements (Green et al., 2007; Whitt, 2007; Li et al., 2014). In experiments, we simulate five demand scenarios: Cycle 1 (UpDown, i.e., demand increasing first but decreasing later), Cycle 2 (DownUp, i.e., demand decreasing first but increasing later), Random Decrease, Random Increase, and Random Fluctuation. In the first two scenarios, the demand will go up and down (Cycle 1) or down and up (Cycle 2), but eventually return to the original level at the end of the planning horizon. In the rest of the scenarios, the demand will randomly decrease (Random Decrease) or increase (Random Increase), or fluctuate around a constant (Random Fluctuation) in time. In each demand scenario, we generate 100 instances and in each instance, we randomly generate the values of $d_t$ in 10 periods following the features of the demand scenario. All instances have $|I| = 4$ and $d_t$ starting with 1000 on average, upper bound 2000, and lower bound 0. The simulation results are summarized below:

| (1) Demand Scenarios | (2) Total Cost (Hierarchical) | (3) Total Cost (Joint) | (4) Total Cost (Integrated) | (5) Cost Saving (J vs. H) | (6) Cost Saving (I vs. H) | (7) Cost Saving (I vs. J) |
|---|---|---|---|---|---|---|
| Cycle 1 (UpDown) | 10,822.22 | 8,703.43 | 8,496.13 | 19.58% | 21.49% | 2.38% |
| Cycle 2 (DownUp) | 5,226.83 | 3,919.93 | 3,877.92 | 25.00% | 25.81% | 1.07% |
| Random Decrease | 4,924.92 | 4,175.06 | 4,096.27 | 15.23% | 16.83% | 1.89% |
| Random Increase | 9,906.07 | 8,330.94 | 8,030.72 | 15.90% | 18.93% | 3.60% |
| Random Fluctuation | 8,944.98 | 7,308.62 | 7,166.07 | 18.29% | 19.89% | 1.95% |

Table 2. Cost Comparison of Decision Approaches in Different Demand Scenarios

Columns 2 to 4 in Table 2 record the average total cost among the 100 instances of each demand scenario, calling each decision approach in one time. Columns 5 to 7 further compare the percentage of cost savings of the joint approach to the hierarchical approach, the integrated approach to the hierarchical approach, and the integrated approach to the joint approach. The results of Table 2 confirm that in all demand scenarios, the both the joint and integrated decision approaches can achieve a significantly lower total cost than the baseline hierarchical decision approach, and that by adopting the integrated approach, the cost can be further reduced. These results provide initial support for the benefit of putting technology planning and workforce planning on the same strategic level and, to a lesser extent, the inclusion of resource allocation at the strategic level in the integrated model. Furthermore, the integrated approach provides the greatest cost saving when the demand generally increases (Random Increase Scenario), or increases first but decreases later (Cycle 1, UpDown Scenario). Note that the above two demand scenarios require a series of capacity expansion decisions. In the other demand scenarios, the demand is decreasing, or not increasing much compared to the initial demand. Thus, either the capacity expansion decisions are not needed, or they incur relatively small costs. However, the demand increase case is most interesting to us because, in this setting, the company has a strong motivation to invest in additional technology and workforce. And in this case, an appropriate decision approach is most needed since the company faces the decision of how acquire and strategically align new technology and workforce resources.

**4.2.2. Decisions in Complex Technology and Workforce Settings**

In the experiments, we gradually increase the size of technology type set *I* from 2 to 10. Consequently, the size of corresponding employee type set *J* increases from 4 to 1024. When the size of the two sets increases, the service company finds a dramatically increased number of resources available in the market, which, therefore, makes the optimal resource planning decisions more difficult. For each size of technology type set *I*, we randomly generate 100 different instances and solve each instance using the models provided by each decision approach. The following table compares the average costs of the three decision approach in each technology and workforce setting; it also records model size, in terms of the number of variables, and computational time for the models of the joint and the integrated approaches.

| Settings | | Total Cost | | | Cost Saving | | | Model Size (No. of Var.) | | Comp. Time (Second) | |
|---|---|---|---|---|---|---|---|---|---|---|---|
| $|I|$ | $|J|$ | Hierarchical | Joint | Integrated | (J vs. H) | (I vs. H) | (J vs. I) | Joint | Integrated | Joint | Integrated |
| 2 | 4 | 9,584.09 | 8,302.29 | 8,108.50 | 13.37% | 15.40% | 2.33% | 60 | 280 | 0.08 | 0.17 |
| 3 | 8 | 9,500.18 | 8,243.97 | 7,984.26 | 13.22% | 15.96% | 3.15% | 90 | 740 | 0.21 | 0.24 |
| 4 | 16 | 9,530.52 | 8,225.05 | 7,945.73 | 13.70% | 16.63% | 3.40% | 120 | 2,160 | 0.16 | 0.78 |
| 5 | 32 | 9,814.40 | 8,330.21 | 7,987.49 | 15.12% | 18.61% | 4.11% | 150 | 6,980 | 0.19 | 1.38 |
| 6 | 64 | 9,883.03 | 8,338.32 | 7,902.71 | 15.63% | 20.04% | 5.22% | 180 | 24,440 | 0.22 | 3.03 |
| 7 | 128 | 9,720.43 | 8,382.47 | 7,997.20 | 13.76% | 17.73% | 4.60% | 210 | 90,380 | 0.32 | 25.13 |
| 8 | 256 | 9,803.24 | 8,422.75 | 8,045.04 | 14.08% | 17.93% | 4.48% | 240 | 345,760 | 0.36 | 150.07 |
| 9 | 512 | 10,334.77 | 8,557.54 | 8,095.79 | 17.20% | 21.66% | 5.40% | 270 | 1,349,300 | 0.45 | 882.45 |
| 10 | 1024 | 10,323.69 | 8,535.38 | 8,174.81 | 17.32% | 20.82% | 4.22% | 300 | 5,325,000 | 1.57 | 5138.18 |

Table 3. Comparison of Decision Approaches in Different Technology and Workforce Settings

When the available types of technology increase linearly from 2 to 10, Table 3 shows that the sizes of the decision models for the joint approach increase linearly while the sizes for the integrated increase exponentially. Therefore, the computational times used to solve the same problem are significantly different for the two decision approaches as the data shown in the last two columns of the table. Meanwhile, in terms of cost saving, Table 3 indicates that the change of the decision approaches from hierarchical to joint helps reduce, on average, 14.82% of the total cost across all instances, while the change from joint to integrated further reduces the total cost by 4.10%. Although the difficulty of solving the integrated model increases significantly in the size of set $I$, the cost savings from joint to integrated also generally increases.

### 4.2.3. Decisions in Different Cost and Time Settings

In the parameter settings discussed in Section 4.1, we let the workforce-related cost be higher than the technology-related cost since this reflects the case in many service companies. For knowledge-intensive services, in particular, the service process tends to involve high wage professionals and as well as significant labor content, because much of the process cannot be automated (Marjanovic and Freeze, 2012; Froehle and White, 2014).

The workforce-related cost consists of four components: hiring, salary per period, training, and firing. We study how the changes of these cost components impact the total cost produced by the three decision approaches. Denote the parameter settings in Section 4.1 as the default setting. Our experiments double or triple the range of each workforce-related cost component in each setting. Each experiment randomly generates 100 instances and each instance has |$I$| =4 and the Random Increase demand pattern. The following table reports the average total cost produced by the three decision approaches in 9 cost settings.

| Cost Setting | Total Cost (Hierarchical) | Total Cost (Joint) | Total Cost (Integrated) | Cost Saving (J vs. H) | Cost Saving (I vs. H) | Cost Saving (I vs. J) |
|---|---|---|---|---|---|---|
| C0: Default Setting | 9,906.07 | 8,330.94 | 8,030.72 | 15.90% | 18.93% | 3.60% |
| C1: Double Hiring | 11,079.69 | 9,100.18 | 8,541.76 | 17.87% | 22.91% | 6.14% |
| C2: Triple Hiring | 12,254.89 | 9,848.80 | 9,023.96 | 19.63% | 26.36% | 8.37% |
| C3: Double Salary | 16,403.90 | 13,309.21 | 13,129.30 | 18.87% | 19.96% | 1.35% |
| C4: Triple Salary | 22,900.10 | 17,877.78 | 17,780.08 | 21.93% | 22.36% | 0.55% |
| C5: Double Training | 10,146.83 | 8,820.15 | 8,559.71 | 13.07% | 15.64% | 2.95% |
| C6: Triple Training | 10,384.64 | 9,284.45 | 9,071.90 | 10.59% | 12.64% | 2.29% |
| C7: Double Firing | 9,906.07 | 8,403.27 | 8,043.40 | 15.17% | 18.80% | 4.28% |
| C8: Triple Firing | 9,906.07 | 8,411.86 | 8,043.40 | 15.08% | 18.80% | 4.38% |

Table 4. Cost Comparison of Decision Approaches in Different Cost Settings

In the 9 cost settings (C0-C8) shown in Table 4, C0 represents the default setting and each of the other settings represents a deviation from the default setting that is related to one of the four cost components. In general, when the value of one cost component increases, the total costs produced by all decision approaches also increase. But, as settings C7 and C8 indicate, a triple increase of the firing cost has little impact on the total cost. We attribute this to the increase demand pattern in the instances when the service company has little motivation to fire employees as its existing capacity is already below the demand and the firing itself becomes too expensive. Instead, the service company will be more inclined to train current employees. However, training also requires additional cost. The total costs in Settings C5 and C6 suggest that when training becomes more expensive, the benefit of adopting the integrated model over the joint model decreases. Moreover, Settings C1 to C4 show that the hiring cost and the salary cost have the opposite impact on the performance of the integrated model. When the increased hiring cost leads to a higher cost saving of the integrated model over the joint model, the increased salary cost reduces savings. Together, the results of Table 4 suggest that the integrated approach can significantly outperform the joint approach in the cases of costs related to high hiring, low salary, low training, and mildly high firing. Among these cases, the high hiring cost seems to impact the cost savings the most.

We also investigate the impact of the training time settings on the total costs produced by the three decision approaches. In the default setting, the time to obtain one skill ranges from 0 to 2, with an average of 1 period. In experiments, we changed the setting by allowing the average training time to be 0, 1, and 2, respectively. The following table shows the results of these experiments.

| Time Setting | Total Cost (Hierarchical) | Total Cost (Joint) | Total Cost (Integrated) | Cost Saving (J vs. H) | Cost Saving (I vs. H) | Cost Saving (I vs. J) |
|---|---|---|---|---|---|---|
| T1: Avg. Training Time =0 | 9,769.35 | 8,225.04 | 7,804.38 | 15.81% | 20.11% | 5.11% |
| T2: Avg. Training Time =1 | 9,906.07 | 8,330.94 | 8,030.72 | 15.90% | 18.93% | 3.60% |
| T3: Avg. Training Time =2 | 10,157.72 | 8,591.44 | 8,419.11 | 15.42% | 17.12% | 2.01% |

Table 5. Cost Comparison of Decision Approaches in Different Time Settings

Table 5 indicates that the cost saving of the integrated approach compared to the joint approach decreases in the increase of the training time. Note that an employee will have zero productivity during the training period. Thus, a long training time makes training a less attractive option to execute and therefore reduces the cost differences between the joint approach and the integrated approach.

### 4.3. Solution Comparison of Decision Approaches

The experiments in Section 4.2 indicate that the total costs in the optimal solutions provided by the three decision approaches are significantly different: the joint and integrated approaches can provide a much lower cost solution than the hierarchical approach, while the integrated approach further outperforms the joint approach. In this section, we analyze the key features of these solutions that drive these cost differences. The analysis is based on the experiment results of instances with $|I|$=4, the Random Increase demand pattern, and the default cost and time settings specified in Section 4.1. Other settings will give similar results.

We first decompose the solutions of the three decision approaches in terms of individual cost components. The results are shown in the following table.

| Decision Approach | Technology-related Cost | | | Workforce-related Cost | | | | Assign. Cost |
|---|---|---|---|---|---|---|---|---|
| | Purchas. | Discard. | Sum | Hiring | Firing | Cross-Train. | Sum | |
| Hierarchical | 592.04 | 0.00 | 592.04 | 7,925.19 | 0.00 | 0.00 | 7,925.19 | 1,388.83 |
| Joint | 665.52 | -16.22 | 649.30 | 6,980.32 | -293.82 | 0.00 | 6,686.51 | 995.13 |
| Integrated | 752.05 | -21.92 | 730.13 | 5,542.37 | -69.98 | 880.82 | 6,353.22 | 947.37 |

Table 6. Statistics of Cost Components of Solutions

Note the value of each cost component shown in Table 6 is the discounted total cost associated with each decision. For example, the purchasing and maintenance cost in the technology-related cost contains both the one time purchasing cost and the discounted total maintenance cost associated with the purchasing decision. This explains why the technology discarding cost and the workforce firing cost are negative as shown in the table because these decisions add to savings for the payment of technology maintenance and employees' salaries. Table 6 shows that the workforce-related cost in all cost components accounts for the largest portion of the total cost, but this value decreases when changing from hierarchical to joint to integrated models. Conversely, the value of technology-related cost gradually increases in the same sequence. This trend suggests that a company will spend more on technology, but less on workforce when adopting the joint or the integrated decision approach, especially in the context of this study where wages are relatively high compared to capital. This is consistent with the unbalanced spending in today's businesses as reported by Rampell (2011). Moreover, the integrated decision incurs a significant amount of cross-training costs, but leads to the lowest total cost. Note that although the hierarchical model allows cross-training in the workforce planning stage in Step 2, an "optimal" solution will not recommend this option because of the lack of strategic integration of technology and workforce planning decisions. We also analyze solutions in terms of the amount of technology, the size of the workforce, and the utilization of the two types of resources. Findings of the analysis are reported below:

| (1) Decision Approach | (2) Init.Value of Technology & Workforce | Technology Related Decision | | Workforce Related Decision | | Average Statistics | | | |
|---|---|---|---|---|---|---|---|---|---|
| | | (3) Amt. Purchased | (4) Amt. Discarded | (5) No. Hired | (6) No. Fired | (7) Amt. of Technology | (8) Size of Workforce | (9) Utilization of Technology | (10) Utilization of Workforce |
| Hierarchical | 2.58 | 2.27 | 0.0 | 2.27 | 0.0 | 4.02 | 4.06 | 97.11% | 96.23% |
| Joint | 2.0 | 4.2 | 2.7 | 4.2 | 2.7 | 3.27 | 3.27 | 86.54% | 86.54% |
| Integrated | 2.0 | 5.04 | 3.83 | 1.75 | 0.61 | 3.25 | 2.79 | 80.30% | 93.65% |

Table 7. Statistics of Decision Values of Solutions

For each decision approach, Table 7 first shows the initial value of technology and workforce, that is, the value of *BX* and *BY* in the models. Because of the balanced assumption discussed in Section 4.1, the total amount of the technology is equal to the total number of employees, initially. Table 7 also shows the hiring and discarding decisions associated with technology planning and the hiring and firing decisions associated with workforce planning. For example, as shown in column 3, Amt. Purchased, the hierarchical decision purchases 2.27 pieces of technology, on average, during the planning horizon, contrary to 4.2 with the joint and 5.04 with the integrated decision. The last four columns of Table 7 record the average amount of technology and the average size of the workforce, and their utilization, respectively. Table 7 uncovers important features of the three decision approaches. First, compared to the hierarchical decision, both the joint and the integrated approaches purchase more new technology and meanwhile discard more existing technology. As a result, both approaches hold a lower amount of inventory of technology than the hierarchical approach as shown in Column 7. Therefore, we conclude that the two approaches encourage more frequent technology updates and tend to adopt more advanced technology than the hierarchical approach. Second, in terms of the alignment of technology and workforce, the decisions of the joint approach must be balanced. If a new piece of technology is purchased, a new employee will also be hired. Consequently, if a piece of technology is discarded, the employee who operates it will also be fired. Therefore, the utilization of the technology and the workforce are always the same. On the contrary, the decisions of the hierarchical approach are "semi-balanced", i.e., if a new piece of technology purchased, a new employee will also be hired simultaneously. However, if a piece of technology is discarded, the employee who operates it may remain in some cases because of an expensive firing cost. As a result, the utilization of the workforce is slightly lower than the technology in the hierarchical approach. Furthermore, the decisions of the integrated approach are highly unbalanced. While purchasing new technology, the integrated approach hires only a few new employees and discards a large portion of the existing technology. These unbalanced decisions also result in the high workforce utilization and the low technology utilization. Third, the hierarchical decision maintains the largest amount of the resources (both technology and workforce) and has the highest resource utilization, yet has the worst performance in terms of the total cost. This suggests that high resource utilization is not always positively associated with good performance.

## 5. Conclusions

### 5.1. Guidelines to Decision Approach Adoption

The hierarchical model has been a popular resource planning approach due to the simplicity of its three easy steps and its focus on the common production priority of high resource utilization, which is consistent with what we found in our analysis. In the traditional manufacturing environment, technology is viewed as the most important asset of a company. Thus, operating the expensive technology at all possible times is a very sensible decision. However, in today's technology-oriented and knowledge-intensive service companies, technology becomes relatively less important and employees must also be acknowledged as a critical asset. By failing to consider technology and workforce at the same strategic level, the hierarchical approach performs poorly and always leads to the highest cost among the three approaches for the following reasons: The hierarchical approach, by continuing to employ out-of-date technology and low-tech labor at nearly 100% utilization, discourages companies from discarding old technology and purchasing new technology, as well as from training employees to use the updated technology. As a result, the hierarchical approach tends to hire too many low-tech employees and retain low-capacity technology. The unit cost of each type of employee or technology is low, but the total cost results are too high.

The numerical experiments also help us understand why the integrated approach always outperforms the joint approach while the joint always outperforms the hierarchical. The hierarchical approach puts the technology as the top priority, therefore ignoring the need for alignment between the technology and the workforce. The joint approach is able to achieve lower cost than the hierarchical approach because it puts both the technology and the workforce as equivalently important resources and finds the best match for them. Such a match, however, is static, i.e., the two types of resources are paired *a priori*. The integrated model further extends the joint model by dynamically matching the resources through cross-training. By cross-training instead of hiring, employees are able to operate new and more advanced technology. Thus, even though the cost of the labor continually increases in today's market, the integrated approach helps companies to offset this impact by using an unbalanced labor and capital spending strategy.

On the other hand, the adoption of the integrated approach faces a number of challenges. Besides the computational and managerial challenges, which are beyond the scope of this paper, the numerical tests show that solutions from the integrated model may not always be dramatically different from those provided by the joint model. This is true especially in one or a combination of the following conditions: decreasing demand, less complex technology and workforce environments, low hiring cost, high salary, high training cost, low firing cost, and long training periods. In these conditions, the joint decision approach can serve as a good substitute for the integrated approach. Since the joint approach does not include cross-training but the integrated approach does, this comparison is suggestive of conditions when cross-training is less attractive. Moreover, all our experiments show that the joint decision approach can reduce at least 10% of the total cost from the hierarchical approach under all conditions. Considering that the joint approach is easy to implement and its decision models are also easy to solve, we recommend that technology-oriented and knowledge-intensive service companies at least adopt the joint approach, and eventually adopt the integrated approach, if possible.

Interestingly, even if companies have not adopted the integrated decision approach, the recent economic data suggests that they are taking actions consistent with this approach by spending more on new and advanced technology, but less on hiring new employees since 2009. "A capital rebound that sharp and a labor rebound that slow have been recorded only once before – after the 1982 recession" (Rampell, 2011). Our results further demonstrate that cross-training is a key to making this a viable resource planning and allocation approach.

## 5.1. Limitations and Opportunities for Future Research

There are a number of possible extensions to this study. Our model considers only the technical skills (i.e., the hard skills) of a workforce but ignores the soft managerial, business, and interpersonal skills. A particular reason for doing this is because an employee is matched with technology mainly based on his or her technical skills. However, a large body of literature has addressed the importance of these soft skills on workforce performance. Additional studies could try to bring these soft skills into the integrated model framework and further model the cooperation and competition of employees. Additionally, as stated in Propositions 1 and 2, we identify the special structure of the models in the condition of non-negative technology discarding cost. A further investigation could be conducted to identify the problem structures, even if the discarding cost is negative. Finally, we incorporate cross-training in our models by including training time, cost, feasible training options, and productivity loss during the training period. Because we need to meet demand in the same period, this limits the amount of cross-training that can occur during the period. An extension would be to include loss of cross-trained employee efficiency compared to specialists.

# Appendices

## Appendix A. The Hierarchical Models

### Step 1. Technology Planning.

$$\textbf{H\_S1}: \qquad HEP(x_{it}^*, v_{it}^*) = \min \ \sum_{t=1}^{T} \gamma^{t-1} \sum_{i \in I} (p_{it} x_{it} + s_{it} v_{it}) \tag{A1}$$

$$s.t. \qquad \sum_{i \in I} c_i \left( BX_i + \sum_{t'=1}^{t} x_{it'} - \sum_{t'=1}^{t} v_{it'} \right) \geq d_t, \quad \forall 1 \leq t \leq T \tag{A2}$$

$$x_{it} \text{ and } v_{it} \text{ are non-negative integers} \tag{A3}$$

Model **H_S1** minimizes total technology-related cost, which consists of technology purchasing, discarding, and maintenance cost. Constraint (A2) ensures that the desired capacity level in each period $t$ must be no larger than the total technology capacity available in that period, which is equal to the capacity of the initial existing technology, plus the increased capacity from purchased technology until $t$, minus the diminished capacity from discarded technology until $t$. Denote $\{x_{it}^*, v_{it}^*\}$ as the optimal technology planning decisions and $HEP(x_{it}^*, v_{it}^*)$ as the minimum cost in **H_S1**.

### Step 2. Workforce Planning.
**H_S2**:

$$HWP(y_{jt}^*, w_{jt}^*, u_{jj't}^* \mid x_{it}^*, v_{it}^*) = \min \ \sum_{t=1}^{T} \gamma^{t-1} \left[ \sum_{j \in J} h_{jt} y_{jt} + \sum_{(j,j') \in \Omega_I} \sum r_{jj't} u_{jj't} + \sum_{j \in J} f_{jt} w_{jt} \right] \tag{A4}$$

$$s.t. \qquad \sum_{\{j \mid (i,j) \in \Psi\}} Q_{ijt} = BX_i + \sum_{t'=1}^{t} x_{it'}^* - \sum_{t'=1}^{t} v_{it'}^*, \quad \forall i, t \tag{A5}$$

$$Y_{jt} \geq \sum_{\{i \mid (i,j) \in \Psi\}} Q_{ijt}, \quad \forall j, t \tag{A6}$$

$$Y_{jt} = BY_j + y_{jt} + \sum_{\{j' \mid (j',j) \in \Omega, t'=t-l_{j'j}>0\}} u_{j'jt'} - \sum_{\{j' \mid (j,j') \in \Omega\}} u_{jj't} - w_{jt}, \quad t=1, \forall j \tag{A7a}$$

$$Y_{jt} = Y_{j,t-1} + y_{jt} + \sum_{\{j' \mid (j',j) \in \Omega, t'=t-l_{j'j}>0\}} u_{j'jt'} - \sum_{\{j' \mid (j,j') \in \Omega\}} u_{jj't} - w_{jt}, \quad 2 \leq t \leq T, \forall j \tag{A7b}$$

$y_{jt}$, $u_{jj't}$, $w_{jt}$, $Y_{jt}$, and $Q_{ijt}$ are non-negative integers (A8)

Model **H_S2** determines the optimal workforce decision given the available types and quantity of technology determined by Step 1. The model introduces two new intermediate variables. Variable $Y_{jt}$ defines total available employees of type $j$ in period $t$, and variable $Q_{ijt}$ defines the planned match between employees and technology in each period. Constraints (A5) and (A6) ensure that each piece of technology must have a qualified employee to operate it at any time and that each employee can work on, at most, one piece of technology at any time. Constraints (A7a and b) specify the balance of workforce flows in each period. The total available employees of type $j$ in period $t$ is equal to the total available employees in the previous period, plus two incoming flows (hired type $j$ employees and transformed employees who have transformed into type $j'$ through training) and minus two outgoing flows (fired type $j$ employees and type $j$ employees who start training to become other types). The objective is to minimize total workforce-related costs, which consist of hiring, firing, training, and salary costs. Denote $\{y_{jt}^*, w_{jt}^*, u_{jj't}^*, Y_{jt}^*, Q_{ijt}^*\}$ as the optimal workforce planning decisions and $HWP(y_{jt}^*, w_{jt}^*, u_{jj't}^* \mid x_{it}^*, v_{it}^*)$ as the minimum cost in **H_S2**.

**Step 3. Resource Assignment.**

**H_S3**: $\quad HRA(z_{ijt}^* \mid y_{jt}^*, w_{jt}^*, u_{jj't}^*, x_{it}^*, v_{it}^*) = \min \ \sum_{t=1}^{T} \gamma^{t-1} \sum_{i \in I} \sum_{\{j \mid (i,j) \in \Psi\}} m_{ijt} z_{ijt}$ (A9)

s.t. $\quad \sum_{i \in I} \sum_{\{j \mid (i,j) \in \Psi\}} c_i z_{ijt} \geq d_t, \quad \forall t$ (A10)

$z_{ijt} \leq Q_{ijt}^* \quad \forall i, j, t$ (A11)

$z_{ijt}$ are non-negative integers (A12)

Model **H_S3** determines the optimal assignments between employees and technology to satisfy the capacity requirements. Note that variable $z_{ijt}$ in the model is different from the intermediate match variable $Q_{ijt}^*$ in model **H_S2**. Here, $z_{ijt}$ specifies the actual assignment between employees and technology based on the particular capacity requirement during each period, while $Q_{ijt}$ represents the match between two types of resources. In other words, a matched resource pair $(i, j)$ with $Q^*_{ijt} = 1$ in period $t$ may not be assigned to serve customers (i.e., $z_{ijt} = 0$) if there is already a sufficient capacity from other assignments to satisfy the capacity requirement. Denote $\{z^*_{ijt}\}$ as the optimal service assignment decisions and $HRA(z^*_{ijt} \mid y^*_{jt}, w^*_{jt}, u^*_{jj't}, x^*_{it}, v^*_{it})$ as the minimum cost in **H_S3**.

**Appendix B: Proof of Proposition 1**.

**Proof**. If $s_{it} \geq 0$ for all $i, t$, it is clear that the optimal solution has $v_{it} = 0$ for all $i, t$. Model **H_S1** is reduced into:

**H_S1'**: 
$$HEP(x^*_{it}) = \min \; \sum_{t=1}^{T} \gamma^{t-1} \sum_{i \in I} (p_{it} x_{it}) \tag{B1}$$

s.t. 
$$\sum_{i \in I} c_i (BX_i + \sum_{t'=1}^{t} x_{it'}) \geq d_t, \quad \forall 1 \leq t \leq T \tag{B2}$$

$$\Delta x_{it} \text{ are non-negative integers} \tag{B3}$$

**H_S1'** is very similar to the Time-Dependent Knapsack Problem (TDKP), which consists of meeting integral demands at distinct time periods at minimum total discounted cost through a selection of items (with integral costs and capacities) from a collection of *N* distinct types of objects. Saniee (1995) first addressed the problem and provided an efficient pseudo-polynomial time solution. The algorithm of Saniee (1995) includes two parts. Part I is to solve the standard Knapsack Problem with demand equal to *D+C*-1, where $D = \max_{1 \leq t \leq T}\{\max\{d_t - d_{t-1}, 0\}\}$, the maximum increment of demands in two periods, and $C = \max_{i \in I}\{c_i\}$, the maximum capacity over all items. Let $M_k$ be the optimal cost to meet *k* units of demand, $1 \leq k \leq D+C-1$. Part II is to construct a network, in which each node $n_{kt}$ represents the case that at least *k* units of extra capacity exist in period *t* to be used at a later time period, and each arc $(n_{kt}, n_{k't+1})$ represents increased capacity with cost equal to $\gamma^t M_{\max\{d_{t+1}-d_t, 0\}+k'-k}$. The network has a source node *S*, which connects to all nodes in period 1, and a sink node *K*, which is connected by all nodes in period *T*. TDKP is equivalent to solve a shortest path problem from *S* to *K* in the constructed network.

Although the similarity, model **H_S1'** has two features different from the classical TDKP: one, **H_S1'** has a set of initial technology, $BX_i$, while TDKP does not, and two, the cost function $p_{it}$ in **H_S1'** is time dependent while $p_i$ in TDKP is not. To deal with the initial technology set, we transfer constraint (B2) into an equivalent form as follows:

$$\sum_{i \in I} \sum_{t'=1}^{t} x_{it'} c_i \geq d_t', \quad \forall 1 \leq t \leq T \tag{B2'}$$

where $d_t' = \max(d_t - \sum_{i \in I} BX_i c_i, 0)$. To deal with the time-dependent purchasing and maintenance cost, we modify Part I of Saniee (1995): Instead of solving one standard Knapsack Problem, we solve *T* standard Knapsack problems with cost function $p_{it}$ in each period *t*. Consequently, $M_{kt}$ is the optimal cost to meet *k* units of demand in period *t*. In Part II of the algorithm, the cost of arc $(n_{kt}, n_{k't+1})$ is also changed into $\gamma^t M_{\max\{d_{t+1}-d_t,0\}+k'-k,t+1}$. All the others are the same as those found in Saniee (1995). Thus, with the above modifications, model **H_S1'** is equivalent to solve the shortest path problem from *S* to *K* in the constructed network. ♦

**Appendix C: Proof of Proposition 2**.

**Proof**. As shown in the proof of Proposition 1, if $s_{it} \geq 0$ for all $i, t$, the optimal solution has $v_{it} = 0$ for all $i, t$; Problem H_S2 has increased capacities with time for all technology types. In addition, by the balance assumption $\sum_{j \in J} BY_j = \sum_{i \in I} BX_i$, and because the technology is never discarded, the initial workforce is always matched with the initial technology. Therefore, we can remove $BX_i$, $BY_j$ and $v_{it}$ from the formulation of H_S2 in the case of $s_{it} \geq 0$. With the simplified formulation, we further show the following property holds: Given $x_{it}^* > 0$ for $i \in I$ in $t \in T$, if an optimal solution matches a $x_{it}^*$ number of qualified employees to operate the technology in time $t$, i.e., $\sum_{(i,j) \in \Psi} Q_{ijt} = x_{it}^*$, then, the decision of matching the same types and number of employees in the rest of the planning horizon (i.e., $\sum_{(i,j) \in \Psi} Q_{ijt'} = x_{it}^*, \forall t' > t$) is also optimal. The property holds because without discarding, an $x_{it}^*$ amount of technology will always exist and require the same number of qualified employees to operate. By keeping the same match, the cost is zero while adjusting the match will always incur additional cost. Thus, the multi-period decision problem of H_S2 is further reduced into a single period decision problem: Given $x_{it}^* > 0$ for $i \in I$ in $t \in T$, find the same number of qualified employees with minimum cost. Because there is no upper bound constraint on the number of employees that can be hired for each type, it is easy to show that the optimal solution must choose a *single* employee type. Therefore, the problem is further reduced into finding a qualified employee type $j$ for each $x_{it}^* > 0$ with minimum cost. To find such a employee type, we can generate a network, in which each node represents a employee type and each arc represents a feasible training option. Note that each arc has two attributes: training and incremental workforce cost and training time. In addition, hiring can be modeled as a special arc with the initial unit workforce cost and zero time. The problem is equivalent to finding the shortest cost path in which the total time is no more than $t$. This is a classical time-constrained shortest path problem (Ahuja et al., 1993). The modeling details of the problem and its solution algorithm can further refer to the Workforce Selection Problem (**WSP**) presented in Section 3.3 of this paper. ♦

**Appendix D: Proof of Proposition 3**.

**Proof**. Note that model **H_S3** can be decomposed into $T$ single-period problems, each corresponding to a bounded Knapsack Problem with the item size $|I|*|J|$ at time $t$. It is well known that a bounded Knapsack Problem can be solved by dynamical programming in $O(|I|*|J|*D^2)$ where $D = \max_{1 \leq t \leq T} d_t$. Martello and Toth (1990) show that the most effective way to solve the problem is to transform it into the 0-1 Knapsack Problem and then apply a much simpler dynamic programming algorithm to solve. We refer the reader to Martello and Toth (1990) for technique details of the algorithm implementation. ♦

**Appendix E: Cost Parameter Configuration in Numerical Experiments and Sample Instance**

We use the following setting in the numerical experiments for the cost components:

- For workforce-related cost, *Hiring* = [500, 2000]; *Salary* = [200, 300]; *Firing* = [200, 500]; *Training* = [10, 500]

- For technology-related cost, *Purchasing* = [100, 600]; *Maintenance*= [10, 10]; *Discarding* = [10, 20]

- The workforce-technology *Assignment* cost = [50, 60]

All costs are randomly generated within the value range specified above and generally comply with the rule that a technology type with a higher capacity has a higher cost than other technology with a lower capacity. Similarly, a more skillful employee type incurs a higher cost than a less skillful type. In addition, the workforce-related cost is more expensive than the technology-related cost since, as previous studies pointed out (e.g., Aksin et al. 2007), the workforce-related cost is high in a typical knowledge-intensive service company. The following table shows a randomly generated sample instance when $|I| = 2$ and $T = 10$.

| Parameters | Values |
|---|---|
| Sets | $K = I = \{i1, i2\}$; $J = \{j0, j1, j2, j12\}$ |
| Demand and Capacity | $d_t = \{1097, 1194, 1298, 1397, 1495, 1594, 1695, 1793, 1898, 1993\}$; $c_i = \{374, 915\}$ |
| Workforce Cost | *Hiring* = $\{875, 1887, 2746, 3758\}$; *Salary* = $\{0, 213, 287, 351\}$; *Firing* = $\{226, 563, 579, 613\}$ |
| Technology Cost | *Purchasing* = $\{154, 470\}$; *Maintenance* = $\{10, 10\}$; *Discarding* = $\{12, 16\}$ |
| Assignment Cost | $m_{ij} = \{ m_{i1,j1} = 50, m_{i1,j12} = 53.2, m_{i2,j2} = 50, m_{i2,j12} = 53.2 \}$ |
| Training Time and Cost | $(Time, Cost)_{jj'} = \{ (0, 97)_{j0,j1}, (1, 437)_{j0,j2}, (1, 437)_{j1,j12}, (0, 97)_{j2,j12} \}$ |
| Others | $T=10$; $\gamma = 0.93$ |

Table E1. A Sample Instance When $|I| = 2$ and $T=10$

This test instance consists of two types of technology and four types of employees. A type *j*0 employee does not have any skills and cannot work on any technology; a type *j*1 employee can operate type $i$1 technology; a type $j$2 employee can operate type $i$2 technology; and a type $j$12 employee can operate both types of technology. Note that the cost corresponding to each decision of the models is time-dependent and can be constructed as the discounted combinations of the cost components shown in the table. For example, if a type $j$1 employee is hired in Period 2, the total initial workforce cost consists of the one-time hiring cost 1887 and the total discounted salaries from Periods 3 to 10, i.e.,

$$\gamma^{t-1} h_{jt} = (0.93)(1887 + 213 \sum_{t'=2}^{t'=10} 0.93^{t'-2}) = 3112.1$$, but if the same type of employee is hired in Period 3,

the total initial workforce cost will be decreased to $\gamma^{t-1} h_{jt} = (0.93)^2 (1887 + 213 \sum_{t'=3}^{t'=10} 0.93^{t'-3}) = 2791.1$.